\documentclass[aps,preprint]{revtex4}
\usepackage[dvips]{graphicx}

\def\la{\langle}
\def\ra{\rangle}

\newcommand{\be}[1]{\begin{equation}\label{#1}}
\newcommand{\ee}{\end{equation}}
\newcommand{\ba}[1]{\begin{eqnarray}\label{#1}}
\newcommand{\ea}{\end{eqnarray}}

\begin{document}
\date{\today}
\title{Non-Hermitian Hamilton operator in open quantum systems
}
\author{Ingrid Rotter}
\affiliation{
Max-Planck-Institut f\"ur Physik komplexer
Systeme, D-01187 Dresden, Germany }


\begin{abstract}
A powerful method for the description of open quantum systems is the Feshbach
projection operator (FPO) technique. In this formalism, the whole function
space is divided into two subspaces that are coupled with one another. One of 
the subspaces contains the  wave
functions  localized in a certain finite region while the  
continuum of extended scattering
wave functions is involved in the other subspace.
The Hamilton operator of the whole system is Hermitian, that of the
localized part is, however, non-Hermitian. 
This non-Hermitian Hamilton operator $H_{\rm eff}$ represents the core 
of the FPO method in present-day studies. It gives a unified description of
discrete and resonance states. Furthermore, it contains the time operator.
The eigenvalues $z_\lambda$ and eigenfunctions $\phi_\lambda$ of 
$H_{\rm eff}$ are an important ingredient of the $S$ matrix. They
are energy dependent. The phases of the $\phi_\lambda$ are, generally, 
nonrigid. 
Most interesting physical effects are caused by the branch points in the
complex plane. On the one hand, they cause the avoided level crossings
that appear as level repulsion or widths bifurcation in approaching the branch
points under different conditions. On the other hand,
observable values are usually enhanced and accelerated 
in the vicinity of the branch points. In most cases, the theory is
time asymmetric. An exception are the 
${\cal PT}$ symmetric bound states in the continuum 
appearing in space symmetric systems due to the avoided level
crossing phenomenon in the complex plane. 
In the paper, the peculiarities of the FPO method are considered
and  three typical phenomena are sketched: (i) the unified
description of decay and scattering processes, (ii) 
the appearance of bound states in the continuum and (iii) the
spectroscopic reordering processes characteristic of the regime with
overlapping resonances.

\end{abstract}

\pacs{03.65.Ca, 03.65.Ta,  03.65.Yz,  05.60.Gg}
 
\maketitle

\section{Introduction}

An exact description of open quantum systems
meets the problem to consider simultaneously
the wave functions of discrete and scattering  states. 
Both types of wave functions are completely different from one another. 
The discrete states $\lambda$ characterize the 
spectrum of the system and are normalized
according to the Kronecker delta $\delta_{\lambda \lambda '}$ 
while the scattering states are
continuous in energy E and can be normalized according to 
the Dirac delta function $\delta(E-E')$.
The wave functions of discrete and scattering states 
appear in a combined manner in most physical expressions characteristic of
open quantum systems. Special mathematical
considerations are necessary therefore in order to receive physical values.

In the $N$-level Friedrichs model \cite{friedrichs,miyamoto}, the total
Hamiltonian $H$ is defined by 
\begin{eqnarray}
H=H_0+\mu V
\label{frie1}
\end{eqnarray}
where $\mu$ is a real number and $H_0$ is the so-called free Hamiltonian
\begin{eqnarray}
H_0= \sum_n \omega_n |n\rangle \langle n |
+ \int_{K_\omega} \omega | \omega\rangle \langle \omega |\, \rho (\omega )
d \omega \; .
\label{frie2}
\end{eqnarray}
Here, $|n \rangle$ and $|\omega\rangle$ satisfy the orthonormality
condition $\langle n | n '\rangle = \delta_{n n '}$
and $\langle \omega | \omega '\rangle = \delta(\omega-\omega ') /
\rho(\omega)$,  and $\langle n | \omega \rangle=0$.
The sum runs over the (finite) number of discrete basic states $|n\rangle$ 
and the integral is over the considered   energy region with 
$K_\omega =\{\omega | \rho (\omega ) \ne 0 \} $. 
The interaction Hamilton operator $V$ describes the coupling
between $|n \rangle$ and $|\omega \rangle$,
\begin{eqnarray}
V= \sum_n \int_{K_\omega} \Big(v_n (\omega) \, 
|\omega\rangle \langle n|
+ v^*_n (\omega) \, | n \rangle\langle \omega | \Big)
~ \rho(\omega) d\omega
\label{frie3}
\end{eqnarray}
where $v_n (\omega)$ is  the interaction matrix element
between $|n \rangle$ and $|\omega \rangle $.
In the Friedrichs model, the Schr\"odinger equation with the Hamiltonian $H$
is directly solved. It is however
not easy to receive results that are of physical interest in a broad range of
parameters.  For an example of the troubles see the study on bound states 
in the continuum  \cite{miyamoto}.

Another method to solve the Schr\"odinger equation with the Hamilton operator
(\ref{frie1}) is the use of the Feshbach projection operator (FPO) technique.
In this method, the basic equations for the wave functions 
of the states $\lambda$ and $\omega$
are solved separately such that the main problem of the Friedrichs model 
is avoided. In the FPO formalism \cite{feshbach}, 
the full function space is divided into two subspaces:
the $Q$ subspace contains all wave functions that are localized inside the 
system and vanish outside of it while
the wave functions of the $P$ subspace are extended up to infinity and vanish
inside the system, see \cite{rep}. 
The wave functions of the two subspaces can be obtained by standard methods: 
the $Q$ subspace is described by the Hermitian Hamilton operator $H_B$
that characterizes the  closed system with discrete states, while
the $P$ subspace  is described by the Hermitian Hamilton
operator $H_C$ that contains the continuum of scattering wave functions.
Thus, $H_0=H_B+H_C$ in (\ref{frie2}). The coupling matrix elements are
calculated according to (\ref{frie3}) by using the eigenfunctions
$\lambda$ of $H_B$ instead of the basic wave functions $n$ 
that appear in (\ref{frie3}). An example for the
difference between the $\lambda$ and the $n$ is the following: in nuclear
structure calculations, the $n$ are determined by the Slater determinants
while the $\lambda$ are the shell-model wave functions.  
Furthermore, $\mu =1$ in (\ref{frie1}), i.e. there is no free parameter in $H$
in the framework of the FPO formalism. All values are determined by fixing 
the potential and the coupling matrix elements between the different 
discrete and scattering states and by 
defining  the two subspaces in a meaningful manner.

In the FPO formalism, the closed system (defined by the 
Hamilton operator $H_B$) will be opened by coupling the wave functions of the 
$Q$ subspace to those of the $P$ subspace under the assumption $P+Q=1$.
Due to this coupling, the discrete states of the closed system 
that lie above particle decay thresholds, pass into
resonance states of the open system. The states below decay thresholds 
receive, as a rule, some energy shift but remain discrete.
The resonance states have, in general, a finite life time. 

The FPO method is introduced by Feshbach \cite{feshbach} forty years ago  
in order to describe nuclear reactions. At that time it was impossible to
perform all the calculations in the two subspaces as well as 
those for the coupling
matrix  $V$. Instead, Feshbach used  statistical assumptions 
for the narrow states of the $Q$ subspace (compound nucleus states)
and treated exactly only the so-called direct (fast) reaction part. 
In this manner it was possible to formulate a unified description of 
nuclear reactions, i.e. of the fast direct nuclear reaction part and 
the much slower compound nuclear reaction part. 
 
In the present-day calculations on the basis of the FPO method, all the 
calculations in the $Q$ subspace are performed with the same accuracy as the
calculations for the corresponding closed system ($Q=1$). Also the coupling
matrix elements $v_\lambda(\omega)$  are calculated.  These calculations 
represent therefore
a unified description of structure and reaction phenomena \cite{rep}. 
They allow to draw 
general conclusions on the behavior of open quantum systems under different
conditions, i.e. by controlling them  in a broad parameter range. 

In the present paper, the FPO method is sketched (Sect. 2) and some 
basic peculiarities are discussed
(Sect. 3). An important feature of the method is the appearance of the
non-Hermitian Hamilton operator $H_{\rm eff}$ in an intermediate stage 
of the model. This Hamilton operator allows a unified description of discrete
and resonance states. The
complex eigenvalues $z_\lambda$ and eigenfunctions $\phi_\lambda$ 
of $H_{\rm eff}$ determine decisively  the $S$ matrix (being unitary
for all parameter values) and observables
as will be shown by means of concrete examples
in the following sections. Typical features of the formalism are  
time asymmetry (Sect. 4), the existence of bound 
(${\cal T}$ symmetric) states in the continuum 
when the system is ${\cal P}$ symmetric (Sect. 5) and spectroscopic reordering
processes  that take place under the influence of branch points in the 
complex energy plane 
in the cross over from the weak-coupling regime to the
strong-coupling one (Sect. 6).

\section{Basic relations  of the Feshbach projection operator
(FPO) formalism}

The basic equation of the FPO formalism  
\begin{eqnarray}
(H-E)\;\Psi^E_C = 0 
\label{Psi}
\end{eqnarray}
is solved in the whole function space by dividing it into the two subspaces
$P$ and $Q$ with $P+Q =1$. The Hamilton operator $H$ is Hermitian.
It contains the decay of the subsystem  localized in the $Q$ subspace, 
into the surrounding $P$ subspace where the decay products can be detected. 
The excitation of the states localized in the $Q$ subspace may take place 
via one of the channels $C$ included in (\ref{Psi}) or by another
process that can be described by a source term $F$ appearing 
on the right-hand side of (\ref{Psi}), for details see  \cite{rede}.

In solving (\ref{Psi})  in the whole function space $P+Q=1$ by using the 
FPO technique, the  non-hermitian Hamilton operator 
\begin{eqnarray}
H_{\rm eff} = H_B + \sum_C V_{BC} \frac{1}{E^+ - H_C} V_{CB} 
\label{heff}
\end{eqnarray}  
appears which contains $H_B$ as well as an additional non-hermitian term 
that describes the coupling of the resonance states via the common environment.
Here  $V_{BC}, ~V_{CB}$ stand for the coupling matrix elements between the 
{\it eigenstates} of $H_B$ and the environment \cite{rep}
that may consist of different continua $C$. 
The operator $H_{\rm eff}$ characterizes the part of the problem that is 
localized in the $Q$ subspace while the operator $H$ describes the problem 
in the whole function space $P+Q$.  Therefore, $H_{\rm eff}$ is non-Hermitian
and $H$ is Hermitian. The operator $H_{\rm eff}$
is explicitly energy dependent and symmetric,
\begin{eqnarray}
(H_{\rm eff} - z_\lambda)\,\phi_\lambda =0 \; ,
\label{phi}
\end{eqnarray}
its eigenvalues $z_\lambda$ and eigenfunctions 
$\phi_\lambda$ are complex. The eigenvalues provide not only the energies 
of the resonance states but also their widths. The eigenfunctions
are biorthogonal. For details see \cite{rep}.

The eigenvalues and eigenfunctions of $H_B$ contain the  
interaction $u$ of the discrete states which is given by the 
nondiagonal matrix elements of $H_B$. This interaction 
is of standard type in closed systems and may be called therefore
internal interaction. The
eigenvalues and eigenfunctions of $H_{\rm eff}$ contain additionally the
interaction $v$ of the resonance states via the 
common continuum ($v$ is used here instead of the concrete 
matrix elements of the second term of $H_{\rm eff}$). 
This part of interaction is, formally, of second order and
may be called  external interaction.
While $u$ and Re$(v)$ cause  level repulsion in energy, 
Im$(v)$ is responsible for the bifurcation of the widths 
of the resonance states  (resonance trapping). 
The phenomenon of widths bifurcation (resonance trapping) 
has been proven experimentally in microwave cavities
\cite{stm}.

Since the effective Hamilton operator (\ref{heff}) depends explicitly  on 
energy $E$, so do its eigenvalues $z_\lambda$ and eigenfunctions
$\phi_\lambda$. Far from thresholds, the energy dependence 
is weak, as a rule, in an energy interval of the order of magnitude of the 
width of the resonance state. 
The solutions of the fixed-point equations
$E_\lambda={\rm Re}(z_\lambda)_{|E=E_\lambda}  $ and of
$\Gamma_\lambda=-2\, {\rm Im}(z_\lambda)_{|E=E_\lambda} $
are numbers that coincide (approximately) with the poles of the $S$ matrix. 
In the FPO formalism, however, it is not necessary to consider the
poles of the $S$ matrix since the spectroscopic information 
on the system follows directly from the complex
eigenvalues $z_\lambda$ and eigenfunctions $\phi_\lambda$  
of $H_{\rm eff}$. Moreover,
in the physical observables related to the  $S$ matrix  
the eigenvalues $z_\lambda$ with their full energy dependence are involved,
see (\ref{smatr}).  Due to this fact, information on  the vicinity
(in energy) of the considered resonance states such as the position of decay
thresholds and of neighboring resonance states is involved in the $S$ matrix
and can be received. Such an information can not be
obtained from the poles of the $S$ matrix being (energy-independent) numbers. 

In contrast to the trajectories $z_\lambda(X)$ of the 
eigenvalues of a Hermitian Hamilton operator
(where $X$ is a certain parameter), those of a
non-Hermitian one may cross in the complex plane.
The crossing points are branch points  (called
exceptional points in the mathematical literature). Physically, they are
responsible for the avoided level crossing phenomenon appearing in their
vicinity. More precisely:
in approaching the branch points under different conditions, we have 
level repulsion (together with widths equilibration) or widths bifurcation
(together with level attraction), see \cite{rep}.  

The eigenfunctions $\phi_\lambda$  of $H_{\rm eff}$
are complex and  biorthogonal.
The normalization condition $\langle \phi_\lambda^{\rm left} |
\phi_\lambda^{\rm right}\rangle =     
\langle\phi_\lambda^*|\phi_{\lambda }
\rangle $ fixes only two of the four free parameters \cite{gurosa}. 
This freedom can be used  in order to provide a smooth transition from 
an open quantum system (with, in general, nonvanishing decay widths
$\Gamma_\lambda$ of its states and biorthogonal wave functions
$\phi_\lambda $) to the corresponding closed one (with 
$\Gamma_\lambda \to 0$ and real wave functions that are normalized
in the standard manner): $\langle\phi_\lambda^*|\phi_{\lambda }
\rangle \to \langle\phi_\lambda|\phi_{\lambda }
\rangle =1 $ if the coupling vectors in the non-Hermitian part of 
(\ref{heff}) vanish.
That means, the orthonormality conditions can be chosen as
\begin{eqnarray} 
\langle\phi_\lambda^*|\phi_{\lambda '}\rangle = \delta_{\lambda, \lambda '} 
\label{biorth1} 
\end{eqnarray}
with the consequence that \cite{rep}
\begin{eqnarray}
\langle\phi_\lambda|\phi_{\lambda}\rangle & \equiv & A_\lambda \ge 1
\label{biorth2a} \\
B_\lambda^{\lambda '} \equiv 
\langle\phi_\lambda|\phi_{\lambda ' \ne \lambda}\rangle & = & -B_{\lambda
'}^\lambda \equiv  
- \,\langle\phi_{\lambda '\ne \lambda }|\phi_{\lambda }\rangle \,; 
\quad |B_\lambda^{\lambda '}| ~\ge ~0  \; .
\label{biorth2b}
\end{eqnarray}
Approaching the branch point where the two eigenvalues $z_\lambda$
and $z_{\lambda '}$ coalesce, $A_\lambda \to \infty$ and 
$|B_\lambda^{\lambda '}| \to \infty$.
The normalization condition (\ref{biorth1}) entails that 
the phases of the eigenfunctions in the overlapping regime
are not rigid: the normalization condition
$\la\phi_\lambda^*|\phi_{\lambda}\ra =1$ is fulfilled, in this regime, 
only when Im$\langle \phi_\lambda^*|\phi_\lambda\rangle \propto $
Re$~\phi_\lambda \cdot$  Im$~\phi_\lambda =0$, i.e. 
by rotating the wave function at a certain angle $\beta_\lambda$. 
For details see \cite{robra}.
The phase rigidity defined by
\begin{eqnarray}
r_\lambda = 
\frac{\langle \phi_\lambda^*| \phi_\lambda \rangle }
{\langle\phi_\lambda | \phi_\lambda \rangle} = 
\frac{1}{({\rm Re}\, \phi_\lambda) ^2 + ({\rm Im}\, \phi_\lambda) ^2}= 
\frac{1}{A_\lambda}
\label{ph2} 
\end{eqnarray}
is a useful measure \cite{brsphas} for the 
rotation angle $\beta_\lambda$. 
When the resonance states are distant from one another, it is 
$r_\lambda \approx 1$ due to $ \langle\phi_\lambda |\phi_\lambda\rangle$ 
$\approx \langle\phi_\lambda^*|\phi_\lambda\rangle$. 
In approaching a branch point in the complex energy plane \cite{rs2,rep}, 
we have $r_\lambda \to 0$. Therefore $1\ge r_\lambda \ge 0$. 

The phase rigidity $r_\lambda$ is a measure for the degree of alignment of 
one of the overlapping resonance states with one of the scattering states 
$\xi^E_C$ of the environment. This alignment takes place at the cost of the
other states that decouple, to a certain extent,  from the 
environment ({\it widths bifurcation} or {\it resonance trapping} \cite{rep}). 
It agrees with experimental data \cite{demb2}, according to which the
phase rigidity drops smoothly from its maximum value
$r_\lambda =1$ far from the branch point 
to its minimum value $r_\lambda =0$ at the branch point, see \cite{robra}.

The solution of (\ref{Psi}) reads \cite{rep}
\begin{equation}
\label{total}
|\Psi_C^E \rangle= |\xi^E_C\rangle + \sum_{\lambda}
|\Omega_\lambda^C \rangle
~\frac{\langle\phi_\lambda^* |V| \xi^E_C\rangle}{E-z_\lambda} 
\end{equation}
where 
\begin{equation}
\label{reswf}
|\Omega_\lambda^C \rangle= 
\Big( 1+ \frac{1}{E^{+}-H_C} V_{CB}\Big)|\phi_\lambda \rangle 
\end{equation}
is the wave function of the resonance state $\lambda$ and
the $\xi^E_C$ are the (coupled) scattering wave functions 
of the continuum into which the system is embedded.
The expression (\ref{total}) follows by applying $P+Q=1$ to (\ref{Psi})
without any approximations. It is therefore an exact solution of (\ref{Psi}). 
The representation of $\Psi^E_C$ (being solution of (\ref{Psi}) with the
Hermitian operator $H$) in the set of wave functions $\phi_\lambda$  
(being solutions of (\ref{phi}) with the non-Hermitian operator $H_{\rm eff}$)
characterizes the consideration of localized states in the FPO formalism.
According to (\ref{total}), the eigenfunctions $\phi_\lambda$ of the 
non-Hermitian Hamilton operator $H_{\rm eff}$ give the main
contribution to the  wave function $\Psi^E_C$ in the interior 
of the system,
\begin{eqnarray}
|\Psi_C^E \rangle \to |\hat \Psi_C^E \rangle =
\sum_\lambda c_{C \lambda}^E\, |\phi_\lambda \rangle  \, ; 
\quad \; c_{C \lambda}^E =
\frac{\langle \phi_\lambda^* |V| \xi^E_C\rangle}{E-z_\lambda } 
\label{total1}
\end{eqnarray}
and 
\begin{eqnarray}
\langle\Psi_C^E | \to \langle\hat \Psi_C^E | =
\sum_\lambda d_{C \lambda}^{E} \, \langle\phi_\lambda^{\rm left} |  
= \sum_\lambda d_{C \lambda}^{E} \, \langle\phi_\lambda^{*} | \; .
\label{total1l}
\end{eqnarray}
The weight factors $c_{C \lambda}^{ E}$ and $d_{C \lambda}^{ E}$ ($=
c_{C \lambda}^{E*} $ in the scattering process) 
contain decay and excitation, respectively, of the states 
$\lambda$ at the energy $E$.

The $S$ matrix is unitary. It can be obtained
from $\langle \xi^E_C|V|\Psi^E_C\rangle $, see \cite{rep}. 
The amplitude of the resonance part we are interested in, is given by
\begin{eqnarray}
S^{\rm res} = i \sum_\lambda \langle\xi^E_C|V|\phi_\lambda\rangle 
~c^E_{C \lambda}
=i\sum_\lambda\frac{ \langle\xi^E_C|V|\phi_\lambda\rangle \langle
  \phi_\lambda^*|V|\xi^E_C\rangle }{E-z_\lambda} \; .
\label{smatr}
\end{eqnarray}
The resonance structure of $S^{\rm res}$ is determined by the eigenvalues
$z_\lambda$ of $H_{\rm eff}$ as long as the resonances do not overlap (see
Sect. 6).

\section{Peculiarities of the Feshbach projection operator
(FPO) formalism. Unified description of discrete and resonance states}

The characteristic features of the FPO formalism consist,
above all,  in the fact that the
solution $\Psi_C^E$ in the whole function space 
can be represented in the set of
wave functions $\{\phi_\lambda\}$ that describe the localized part of the
problem. The localized wave functions represent a subspace of the
whole function space with the consequence that the corresponding 
Hamilton operator $H_{\rm  eff}$ is non-Hermitian.  
 
The main advantages of the FPO formalism consist in the following.

(i) The spectroscopic information on the resonance states is
obtained directly from the complex eigenvalues $z_\lambda$
and eigenfunctions $\phi_\lambda$
of the non-Hermitian Hamilton operator $H_{\rm eff}$.
The $z_\lambda$ and $\phi_\lambda$
are energy dependent functions, generally, and contain the
influence of neighboring resonance states as well as of decay thresholds
onto the considered state $\lambda$. This energy dependence allows to describe 
decay and resonance phenomena also in the very neighborhood of
decay thresholds and in the regime of overlapping resonances.
Since also the coupling coefficients between system and continuum 
depend on energy,  the unitarity of the $S$ matrix is guaranteed
for all parameter values, see e.g. \cite{ro03}.

(ii) The resonance states are directly related to the discrete states 
of a closed system described by standard quantum mechanics (with the Hermitian
Hamilton operator $H_B$). They  are generated by opening the system, i.e. by
coupling the discrete states to the environment of scattering states by 
means of the second term of
the Hamilton operator $H_{\rm eff}$. Therefore, they are realistic
localized (long-lived many-particle)  states of an  open quantum system.

(iii) The properties of branch points 
and their vicinity can be studied relatively easy. At these  points,
two (or more) eigenvalues $z_\lambda$ of $H_{\rm eff}$ coalesce. 
Since it is not necessary to consider the poles of the $S$ matrix
in the FPO formalism, 
additional mathematical problems at and in the vicinity of branch points 
(exceptional points) in the complex plane are avoided.

(iv) The phases of the eigenfunctions $\phi_\lambda$ 
of $H_{\rm eff}$ are not rigid in the vicinity of a branch point.
This fact allows to describe the
spectroscopic reordering processes in the system that take place  
under the influence of the scattering wave functions $\xi^E_C$
of the environment into which the system is embedded.

These features are involved in all present-day \cite{presentday} 
calculations performed on the 
basis of the FPO formalism. In numerical studies, 
the main problem arises from the definition of the
two subspaces $Q$ and $P$ such that it is meaningful for spectroscopic
studies (the cross section is, of course, independent of 
the manner the two subspaces are defined, for more details see  \cite{rep}).
The basic idea is the following: $H_B$ describes the closed system
(localized in the interior of the system)
which becomes open when embedded in the environment of scattering wave
functions described by $H_C$. Therefore, all values characteristic of
resonance states can be traced back
to the corresponding values of discrete states
by controlling the coupling to the continuum. That means with $v\to 0$, 
the transition from resonance states (described by the non-Hermitian 
$H_{\rm eff}$) to discrete states (described by the Hermitian $H_B$)
can be controlled.  

As to the mathematical
properties of branch points in the complex energy plane, 
it will be mentioned here only that the phase jump behavior 
of the wave functions at and in the
vicinity of the branch points and their topological structure are  
investigated in the framework of the FPO formalism by means of 
numerical calculations for some special cases as well as 
in analytical studies \cite{marost23,ro01,rs1,rs2}. 
The topological structure of the branch (exceptional) points in the continuum
differs from that of diabolic points.
The results obtained in the FPO formalism 
are in full agreement with those of rigorous mathematical studies
for a symmetric non-Hermitian 2x2 matrix Hamiltonian 
\cite{gurosa}. At the exceptional 
(branch) point,  the two different right and left 
eigenvectors of the non-Hermitian Hamilton operator  are linearly dependent,
$|\phi_{\lambda }\rangle \leftrightarrow 
\pm ~i |\phi_{\lambda '}\rangle$;
$\langle\phi_{\lambda }^*| \leftrightarrow 
\mp ~i \langle\phi_{\lambda '}^*|$, 
and are supplemented by the corresponding
associated vectors defined by Jordan chain relations.  Moreover, the
results are  in agreement with those of experimental studies on
microwave cavities \cite{demb1}, see Ref. \cite{robra}. 
The cross section ($S$ matrix) behaves
smoothly at the branch point \cite{mudiisro,rep}.

The physical meaning of the branch points in the complex energy plane is
based upon their topological structure 
and their relation to the phenomenon of avoided level crossing, i.e. to, 
respectively, 
level repulsion and widths bifurcation occurring in approaching them under 
different conditions. Level repulsion is accompanied by widths 
equilibration, while widths bifurcation occurs together with level attraction 
(formation of clusters). For further details see e.g.  \cite{rep}.

In the FPO formalism, a unified description of discrete and resonance states is
involved. First, the resonance states pass smoothly into discrete states by
reducing the coupling strength
$v$ between system and environment. The smooth transition
is guaranteed by choosing the orthonormality conditions for the eigenfunctions
$\phi_\lambda$ of $H_{\rm eff}$ according to (\ref{biorth1}). 
Secondly, the localized  states in the $Q$ subspace  are
eigenstates of $H_{\rm eff}$. Generally, they may lie above as
well as below the particle decay thresholds. In the first case, their widths
$\Gamma_\lambda$ are different from zero (with the exception of
bound states in the continuum, see Sect. 5) while they are zero in
the second case. In the last case, $H_{\rm eff}$ is real: the 
residuum of the second term of (\ref{heff}) vanishes and only the
principal value integral is different from zero. It causes, generally, 
some shift of
the position of the state $\lambda$, $\Delta (E^B_\lambda - E_\lambda) \ne 0$,
see \cite{rep}.
Also the wave functions of these states differ, generally, from those of the
eigenfunctions of $H_B$, $\phi_\lambda \ne \phi_\lambda^B$. That means, 
discrete and resonance states are described in a unified manner by
the non-Hermitian Hamilton operator $H_{\rm eff}$:
not only the resonance states lying above particle decay thresholds, but
also the discrete states below the thresholds are influenced by the continuum.
As a consequence, the branch points in the complex plane determine also the
properties of bound states. The most impressive example is the phenomenon of
avoided crossing of discrete states known since the very beginning 
of quantum physics 
\cite{neuwi2}. It can be traced from discrete states up to the branch
points in the complex plane by parameter variation \cite{ro01}.
At the branch points,  nonlinearities play a role. The branch points 
introduce therefore nonlinear effects into quantum mechanics \cite{rep}. 
This fact may lead to a deeper  understanding of
the relation between avoided level crossings and quantum chaos.

These results show that 
the non-Hermitian Hamilton operator $H_{\rm eff}$ is
basic for the  description of the localized states in 
realistic quantum systems. It
describes the spectroscopic properties of discrete and resonance
states in a unified manner.  
The restriction to the Hermitian Hamilton operator
$H_B$ [see (\ref{heff})] in the standard quantum mechanics is an 
approximation: the principal value integral of the second term of (\ref{heff})
is effectively  taken into account, however the residuum is  
neglected. Although this approximation works well in 
{\it very} many applications, it leaves open some fundamental questions.  
In these cases, the non-Hermiticity of the Hamilton operator 
$H_{\rm eff}$ (and the nonlinear effects related to it)  
can not be neglected.

Another peculiarity of the FPO formalism is the existence of
a time operator which is the residuum  of the non-Hermitian Hamilton
operator $H_{\rm eff}$. The life time $\tau_\lambda$ of a resonance state
follows from the eigenvalue $z_\lambda$ of 
$H_{\rm eff}$ in the same manner as the energy $E_\lambda$ of
this state. Both values are fundamentally different from the time $t$ and the
energy $E$. They characterize the state $\lambda$ while $t$ and
$E$ appear as general parameters. In the closed system with the 
Hermitian Hamilton operator $H_B$, only the energies $E_B$ of the states can be
determined. The eigenvalues are real and the widths are zero, $\Gamma_B =0$.
Due to the coupling to the continuum, energy shifts 
$E_\lambda - E_B$ of the states
appear as well as the finite life times $\tau_\lambda
\propto (\Gamma_\lambda - \Gamma_B)^{-1} = 
\Gamma_\lambda^{-1} $ of the resonance states. 
Both, the energy shifts and the finite life times, 
follow from the non-Hermitian
coupling term of $H_{\rm eff}$ [see Eq. (\ref{heff})].
Usually, the numbers $E_\lambda $  and $\Gamma_\lambda$
can  be obtained directly from the $z_\lambda$.
Only in the case the $z_\lambda $ are strongly dependent on energy,
the corresponding fixed-point equations have to be solved.
The energies $E_\lambda$ and  life times $\tau_\lambda$ of the resonance 
states $\lambda$ of an open quantum system are bounded from below
(see \cite{robra} for the discussion of the
brachistochrone problem in open quantum systems). 
Mathematically, the existence of the time operator entails
the time asymmetry involved in the FPO formalism.

\section{Unified description of resonance and decay phenomena}

The time dependent Schr\"odinger equation reads
\begin{eqnarray}
H_{\rm eff} ~\hat\Psi^E(t)=i~\hbar ~\frac{\partial}{\partial t}
~\hat\Psi^E(t) \, .
\label{tdse0}
\end{eqnarray}
The right solutions may be represented, according to (\ref{total1}), 
by an ensemble of resonance states $\lambda$ that describes the decay
of the localized part of the system  at the energy $E$, 
\begin{eqnarray} 
|\hat\Psi^{E~\rm (right)}(t)\rangle & =& 
e^{-iH_{\rm eff} \, t/\hbar} ~|\hat\Psi^{E~\rm (right)}(t_0)\rangle 
\nonumber \\
&=& \sum_{\lambda } ~e^{-i z_{\lambda } \, t/\hbar}  
~c_{\lambda 0}
~|\phi_{\lambda }^{\rm (right)}\rangle 
\label{tdse1}
\end{eqnarray}
with $|\phi_{\lambda }^{\rm (right)}\rangle =|\phi_{\lambda }\rangle$
and  $c_{\lambda 0} = 
\langle\phi_\lambda^* |V|\xi^E_C\rangle/(E-z_\lambda) $.
The $z_\lambda$ and $\phi_\lambda$ are the (energy dependent) 
eigenvalues and eigenfunctions of
the time-independent Hamilton operator $H_{\rm eff}$, Eq. (\ref{heff}),
while the $\xi^E_C$ are the scattering wave functions of the environment.
The left  solution  of (\ref{tdse0}) reads
\begin{eqnarray} 
\langle \hat\Psi^{E~\rm (left)}(t)| & = & 
\langle\hat\Psi^{E~\rm (left)}(t_0)|  ~e^{iH_{\rm eff}^{\dagger} \, t/\hbar} 
\nonumber  \\
& = & \sum_{\lambda } 
~\langle\phi_{\lambda }^{\rm (left)}|  
~d_{\lambda t} ~e^{i z_{\lambda }^* \, t/\hbar} 
\label{tdse2}
\end{eqnarray}
with $\langle\phi_{\lambda }^{\rm (left)}| =\langle\phi_{\lambda }^{*}|$  and
$d_{\lambda t}  =c_{\lambda 0}^{*}
=\langle \xi^E_C|V|\phi_\lambda\rangle/(E-z_\lambda^*)$
for the scattering process. 
It describes the excitation of the system at the energy $E$. 
For other excitation processes, e.g.
 via a source term $F$ on the right-hand side of
(\ref{Psi}), see \cite{rede}. Here $d_{\lambda t} $ is, generally, time
dependent. 

By means of (\ref{tdse1}) and (\ref{tdse2}) the population probability 
\begin{eqnarray} 
\langle \hat\Psi^{E~\rm (left)}_C(t)|\hat\Psi^{E~\rm (right)}_C(t)\rangle = 
\sum_\lambda
c_{\lambda 0} ~d_{\lambda t} ~e^{- \Gamma_{\lambda } t/\hbar} 
\label{tdse3}
\end{eqnarray}
at the energy $E$ can be defined.
The decay rate reads 
\begin{eqnarray} 
k_{\rm gr}(t)& =& - \frac{\partial}{\partial t} ~{\rm ln} 
\, \langle \hat\Psi^{E~\rm (left)}_C(t) | \hat\Psi^{E~\rm (right)}_C(t) \rangle
\nonumber \\
& =& \frac{1}{\hbar}~\frac{\sum_\lambda \Gamma_\lambda 
~c_{\lambda 0} ~d_{\lambda t}
~e^{- \Gamma_{\lambda } t/\hbar}}{\sum_\lambda 
c_{\lambda 0} ~d_{\lambda t} ~e^{- \Gamma_{\lambda } t/\hbar}} \; .
\label{tdse4}
\end{eqnarray}
For an isolated  resonance state $\lambda$,  (\ref{tdse4}) 
passes into the standard expression
\begin{eqnarray}
k_{\rm gr}(t) ~\to ~k_\lambda ~= ~\Gamma_{\lambda }/\hbar 
\; .
\label{kiso}
\end{eqnarray}
In this case, the value $k_\lambda$ is constant in time and  corresponds to 
the standard relation $\tau_\lambda = \hbar / \Gamma_\lambda$
with $\tau_\lambda = 1/k_\lambda $.
It describes the idealized case of an exponential decay law and,
according to (\ref{smatr}),
a Breit-Wigner resonance in the cross section.
Generally, deviations from the exponential decay law and from 
the Breit-Wigner line shape appear under the influence of neighboring
resonance states and (or) of decay thresholds. 
Also the background term appearing in most reactions may cause deviations
from the ideal exponential decay law. For details see \cite{rede}

The expressions (\ref{tdse1}) and (\ref{tdse2}) are valid only  when   
(\ref{total1}) holds, i.e. at  times $t$ at which the 
wave functions $\Psi^E_C $  have a localized part  
in the interior of the system at the energy $E$ so that the representation
(\ref{total1}) is meaningful at this energy. According to (\ref{tdse1})
and (\ref{tdse2}), this is the case for times
$t\ge t_0$ where $t_0$  is  a finite
value. Without loss of generality, it can be chosen $t_0 =0$. 
The quantum system described in the framework of the FPO formalism is
therefore time asymmetric. The time asymmetry is involved in 
the non-Hermitian part of the Hamilton operator $H_{\rm eff}$
(which contains the time operator), as can be seen immediately from the
expression (\ref{tdse3}) for the population probability.

The consideration of only the time interval $0\le t \le \infty$ in 
(\ref{tdse0}) is related to the fact that the decay of a resonance state
(at the energy $E$ of the system) starts at a finite time 
(say $t_0=0$) at which the system
can be considered to be excited, i.e. (\ref{total1}) is meaningful
at this energy. This fact agrees with the concept of a semigroup description 
introduced in \cite{bohm},  
which distinguishes between prepared and measured states.  
In our formalism, the decaying (measured) states are described by the  
eigenvalues and eigenfunctions of the effective non-Hermitian 
Hamilton operator $H_{\rm eff}$ involved in the $|\hat\Psi^E \rangle $,
Eq. (\ref{total1}). 
The preparation of the resonance states is described by the 
energy-dependent $\langle \hat \Psi^E |$.
It may be very different for different reactions. 

The decay properties of the resonance states
can be studied best when their excitation 
takes place in a time interval that is very short 
as compared to the life time $\tau_\lambda$ of the resonance states,
i.e. $d_{\lambda t}$ is a function being
strongly dependent on time. In such a
case, the time $t_0=0$ is well defined and no perturbation of the decay process
by the still continuing excitation process will take place. 
In \cite{bohm2}, such a situation is studied in single ion experiments.
The results demonstrate  the beginning of time for a decaying state. 
That means, they prove the time asymmetry in quantum physics.

Eq. (\ref{tdse4}) describes the decay rate also in the regime of overlapping
resonances, see \cite{decayrate,rede}. The overlapping and
mutual influence of resonance states is maximal at the branch points in
the complex plane where two  eigenvalues $z_{\lambda}$ and
$z_{\lambda '}$ of the effective Hamilton operator $H_{\rm eff}$ coalesce.
Nevertheless, the decay rate is everywhere smooth as
can be seen also directly from (\ref{tdse4}).
This result coincides with the general statement 
according to which all observable
quantities behave smoothly at  singular points. 

Another interesting problem is the saturation of the average decay rate 
$k_{\rm av}$ in the regime of 
strongly overlapping resonances. According to the bottle-neck picture of the
transition state theory, it starts at a certain critical value of
bound-continuum coupling \cite{miller}. This saturation is caused by
widths bifurcation  (formation of long-lived resonance states by 
resonance trapping by a few short-lived  states \cite{rep}) occurring
in the neighborhood of the branch points in
the complex plane \cite{comment,robra}. Widths bifurcation creates long-lived 
resonance states together with a few short-lived resonance states.
The definition of an average life time 
of the resonance states is meaningful therefore only 
for either the long-lived states or the short-lived ones. 
The long-lived (trapped) resonance states are almost decoupled from the
continuum of decay channels. Their  widths $\Gamma_\lambda$ 
saturate with increasing bound-continuum coupling.
The $\Gamma_\lambda$ are almost the same for all  the different states 
$\lambda$, see \cite{rep}, i.e.
$\Gamma_{\rm av} \approx \Gamma_\lambda$ for all long-lived trapped
resonance states. It follows therefore 
\begin{eqnarray}
k_{\rm av} 
\approx \Gamma_{\rm av} /\hbar 
\label{kapav}
\end{eqnarray}
from (\ref{tdse4}). According to the average width $\Gamma_{\rm av}$,
the average life time  of the long-lived states can be defined by 
$\tau_{\rm av} = 1/k_{\rm av} $. Then (\ref{kapav}) is equivalent 
to $\tau_{\rm av} = \hbar / \Gamma_{\rm av}$.
That means, the basic relation
between life times and decay widths of resonance states holds 
not only for isolated resonance states [see Eq. (\ref{kiso})], but
also for narrow resonance states superposed by a smooth background
(that may originate from a few short-lived resonance states 
\cite{rep,brscorr}).
In the last case, the relation holds for the average values 
$\Gamma_{\rm av}$ and $\tau_{\rm av}$. 

The expressions (\ref{smatr}) for the $S$ matrix
and (\ref{tdse4}) for the decay rate
show immediately that the resonance phenomena (described 
by the $S$ matrix) are determined by the decay properties of the resonance 
states (described by the complex eigenvalues $z_\lambda$ 
and eigenfunctions $\phi_\lambda$ of the
non-Hermitian Hamilton operator $H_{\rm eff}$). Thus, the FPO formalism 
provides a unified description of resonance and decay phenomena. 
The expression (\ref{smatr}) shows however also that, generally, 
the energy dependence of the
eigenvalues $z_\lambda$ and eigenfunctions $\phi_\lambda$  of $H_{\rm eff}$
causes deviations from the 
Breit-Wigner resonance shape and the exponential decay law. The deviations
become important for isolated resonance states due to the fact that the decay
thresholds lie at a finite energy \cite{cusp}. 
They appear  mainly in the long-time scale.
This result agrees qualitatively with experimental data \cite{rothe}.
At high level density, deviations appear even in the short-time 
scale  due to the mutual influence of neighbored resonance states, 
see  Sect. 5 and \cite{robra}.

\section{Bound states in the continuum (BICs)}

The question whether or not bound states in the 
continuum (BICs) exist in realistic quantum systems is  of
principal interest and might be as well of interest for applications.
The reason for this interest arises from the fact that the 
system is stabilized at a BIC as well as in its vicinity, and that the
wave function  is localized at all times
inside the system in spite of embedding
it into the continuum of extended wave functions.

Mathematically, the existence of bound states in the continuum
is shown already in 1929 by 
von Neumann and Wigner  \cite{neuwi1}.
In 1985 Friedrich and Wintgen \cite{friewi} considered the problem 
by using the FPO technique. They related the existence of BICs to
avoided level crossings being another quantum mechanical phenomenon 
discussed  by von Neumann and Wigner \cite{neuwi2}  in 1929. 
As discussed in Sects. 2 and 3,
avoided level crossings are caused by branch points and appear in their
vicinity. 

Since BICs are states that do not decay, the population probability
of these states is constant  in time. This fact is called population trapping 
in studies on laser induced continuum structures in atoms
\cite{laser}. Similar results are obtained \cite{marost23} 
in the time independent approach by using the FPO technique
and demanding a vanishing decay width for the BIC. 
In these papers, 
the relation between BICs and the avoided level crossing phenomenon as well as 
the stabilization of the system in a broad range of the parameter values
(characteristic of the laser) is shown explicitly.
A similar study is performed for the transmission through a quantum billiard
where BICs appear at those energies at which the
resonant transmission crosses  a transmission zero \cite{rs3,bicring}. 
Common to all these studies is the definition of a
BIC as a resonance state with vanishing width, 
\begin{eqnarray}
\Gamma_{\lambda_0}|_{(E=E_{\lambda_0})} = 0 \; .
\label{bicdef}
\end{eqnarray}
In the FPO formalism, its energy is obtained from the solution of the 
fixed-point equation  $E_{\lambda_0}=E_\lambda |_{(E=E_{\lambda_0})}$.

Generally, the relation between $\Gamma_\lambda$ and the coupling matrix 
elements is \cite{rep}
\begin{eqnarray}
\Gamma_\lambda =-2\, {\rm Im}(z_\lambda ) 
\le {2\pi} \sum_C |\langle\phi^*_\lambda |V|\xi^E_C\rangle|^2 \; .
\label{h3}
\end{eqnarray}
This expression holds true at all energies. The inequality 
in (\ref{h3}) is caused by the biorthogonality of the functions $\phi_\lambda$
being eigenfunctions of the non-Hermitian operator (\ref{heff}).
As a consequence, a state being decoupled from 
all channels $C$ of the continuum  according to
\begin{eqnarray}
\langle\xi^E_{C}|V|\phi_{\lambda_0}\rangle \to 0
\label{h4}
\end{eqnarray}
is a BIC with $\Gamma_{\lambda_0}\equiv - 2 \, {\rm Im}(z_{\lambda_0}) \to 0$
[the condition (\ref{h4}) is equivalent to 
$\langle \phi_{\lambda_0}^*|V|\xi^E_C\rangle \to 0$ due to the 
symmetry of $H_{\rm eff}$ and the biorthogonality of the $\phi_\lambda$].
The opposite case follows by considering the $S$ matrix, see (\ref{smatr}) for
the amplitude of its resonance part.   
At the position of a BIC, we have $E- z_{\lambda_0} \to 0 $ and, due
to the unitarity of the $S$ matrix, it follows (\ref{h4}) for all $C$.
That means: the decoupling from all channels of 
the continuum described by (\ref{h4})
is a necessary and sufficient condition for a resonance state to be a BIC,
i.e. a state with vanishing decay width $\Gamma_{\lambda_0}=0$.
The wave function of such a BIC is, according to (\ref{reswf}), 
eigenfunction of $H_{\rm eff}$ and, consequently,  localized.

In \cite{biccomm}, the advantage of the FPO method as compared to the
$N$-level Friedrichs model \cite{miyamoto}
in studying  BICs for unstable multilevel systems
is discussed. In contrast to the coupling matrix elements
(\ref{h4}), the form factors 
$\langle\xi^E_{C}|V|\phi_{n_0}^B\rangle$ and 
$\langle \phi_{n_0}^B|V|\xi^E_C\rangle$ considered in \cite{miyamoto} 
contain the basic wave
functions $\phi_{n}^B$ of the Hamiltonian 
$H_B$ of the closed system. 
Since the eigenfunctions $\phi_\lambda$ of $H_{\rm eff}$ can be represented as 
$\phi_\lambda=\sum a_{\lambda,\lambda '}\phi_{\lambda'}^B $
with complex coefficients $a_{\lambda,\lambda '}$
[and $\phi_\lambda^B=\sum b_{\lambda n}\phi_n^B$ with real $b_{\lambda n}$
and the basic wave functions $\phi_n^B$ of discrete states defining $H_0$
according to (\ref{frie2})], a sum of individual 
form factors vanishes at the position of a BIC according to (\ref{h4}),
\begin{eqnarray}
 \sum_{\lambda '} a_{\lambda_0,\lambda '}
\langle\xi^E_{C}|V|\phi_{\lambda '}^B\rangle \to 0   
\, .
\label{h7}
\end{eqnarray}
This equation  is nonlinear since the coefficients  
$a_{\lambda_0,\lambda '}$ depend on the coupling strength. It is therefore 
difficult to obtain a general solution for the multilevel case
in the framework of the Friedrichs model.  

By using the FPO technique
and demanding a vanishing decay width $\Gamma_{\lambda_0}$,
the multilevel problem can be solved. Examples are given in \cite{marost23} for
atoms and in \cite{rs3} for quantum dots.
At sufficiently small coupling strength $v$, 
the condition $\Gamma_{\lambda_0} =0$ can be fulfilled only
when the spectrum of the closed system (described by $H_B$) 
is degenerated. Concrete examples of BICs investigated recently
analytically with the postulation $\Gamma_{\lambda_0} =0$
for a BIC are studies on open quantum billiards with variable shape
\cite{bicring}. The relation of the BICs
to the avoided level crossing phenomenon can be
seen in all examples. The scattering phase jumps by $\pi$
in approaching the BIC  in spite of the fact that no resonance can be seen in
the cross section \cite{rs3}. 

By means of the complex eigenvalues $z_\lambda$ of $H_{\rm eff}$, 
the appearance 
of a BIC can be traced as a function of a certain control parameter $X$,
i.e. by controlling the trajectories $E_\lambda (X)$ and $\Gamma_\lambda (X)$. 
The BIC appears at the point $X=X_0$ where  $\Gamma_\lambda(X_0)=0$.  
It is even
possible to consider the vicinity of the BIC including the cases when 
$\Gamma_\lambda (X')$  is always different from zero and 
$\Gamma_\lambda (X'_0)$ corresponds to the minimum of $\Gamma_\lambda (X')$
with  a small but nonvanishing value $\Gamma_\lambda(X'_0)  \approx 0$.
This feature  of the FPO technique is
invaluable for  applications since the stabilization of the system 
(caused by the vanishing width $\Gamma_\lambda$) must be known
not only at the single point $X_0$  
but also in its vicinity (where $\Gamma_\lambda > 0$, but small)
in order to estimate the possibility of an experimental observation.

Examples of $\Gamma_\lambda (X)$ trajectories 
with  $\Gamma_\lambda (X_0)=0$ as well as with $\Gamma_\lambda (X'_0) 
\approx 0$
are studied on the basis of the FPO method for concrete systems, see 
\cite{marost23} for atoms and \cite{rs3} for quantum dots.
In both cases, the trajectories depend strongly on energy near $X_0$ and 
$X'_0$, respectively. This energy dependence is caused by  
widths bifurcation, i.e. by the avoided level crossing phenomenon.
The interaction of the resonance states via the continuum
(described by the complex non-diagonal matrix elements of the second 
term of $H_{\rm eff}$) plays an important role. The interplay between 
their real and imaginary parts makes possible the appearance of BICs at 
finite (physical) values of the coupling strength \cite{marost23}.
Every BIC appears together with at least one 
other state whose width is enhanced (due to the widths bifurcation)
around $X=X_0$ and whose energy is,
generally, close to $ E_{\lambda_0}$. In the considered cases,
the condition for the exact appearance of a (${\cal T}$ symmetric)
BIC is space reflection symmetry (${\cal P}$ symmetry) of the
system. Violation
of ${\cal P}$ symmetry leads to $\Gamma_\lambda(X'_0) $ small but different
from zero, i.e. to violation of time reflection symmetry  
(violation of ${\cal T}$ symmetry).

The BICs, being   ${\cal PT}$ symmetric states,   
are the result of  widths bifurcation, i.e. finally  
of the existence of branch points in the complex plane. 
The branch points appear as families of trajectories 
when considered as a function of a  parameter, 
as shown in numerical studies for double quantum dots \cite{rs2}. 
As a consequence, also the BICs  appear 
as families of  trajectories by controlling the system by
means of parameters. Besides the 
trajectories of the ${\cal PT}$ symmetric BICs there are always  
trajectories of
states whose widths are large (widths bifurcation). These states
are ${\cal P}$ symmetric, however  
${\cal T}$  symmetry is broken as discussed in Sects. 3 and 4.

It is interesting to remark that, using the condition of 
${\cal PT}$ symmetry, Bender et al. obtained classes of non-Hermitian 
Hamilton operators whose spectra are  real and positive \cite{bender1}. 
Recently, these solutions and their consequences for physical processes 
are discussed intensively in
the literature, see e.g. the 2007  Workshop on Pseudo Hermitian Hamiltonians in
Quantum Physics \cite{jpaproc}.

\section{Spectroscopic reordering phenomena in the 
  overlapping regime}

An unsolved problem in standard quantum mechanics is the 
description of the crossover from the
regime with weak coupling between discrete and continuous states  to 
that with strong coupling between them. For example, an interpolation
procedure between the limiting cases with isolated resonances
at low level density and narrow resonances at high level density 
(superposed by a smooth background term) is introduced in \cite{shapiro}
for the transmission through a quantum dot. In contrast to such an
interpolation procedure, 
the crossover can be described in the FPO formalism by means of 
the  phase rigidity $r_\lambda$ that is reduced 
for many states  $\lambda$ in this regime. 

Let us consider the one-channel case, $C=1$, and $\Psi^E_C \to 
\hat\Psi^E$
in the interior of the system. The right and left wave functions 
follow from 
(\ref{total1}) and (\ref{total1l}) with $d_{\lambda}^E = c_\lambda^{E*}$
when excitation and decay of the state $\lambda$ 
occur via the same mechanism. Therefore the $\hat\Psi^{E}$ 
can be normalized,
\begin{eqnarray}
\la\hat\Psi^{E, {\rm left}} | \hat\Psi^{E, {\rm right}}\ra & = 
& \sum_{\lambda \lambda '}
c_{\lambda}^{E*} c_{\lambda '}^E \; \la\phi_\lambda^* |\phi_{\lambda '}\ra
= \sum_\lambda |c_{\lambda}^E|^2 \equiv 1 \; .
\label{r4}
\end{eqnarray}
The normalization has to be done separately
at every energy $E$ due to the 
explicit energy dependence of the  $c_{\lambda}^E$. 
Moreover,
\begin{eqnarray}
\la\hat\Psi^{E, {\rm left}\,*} | \hat\Psi^{E, {\rm right}}\ra
&=& \sum_{\lambda \lambda '}
c_{\lambda}^E c_{\lambda '}^E \; \la\phi_\lambda |\phi_{\lambda '}\ra
= \sum_\lambda (c_{\lambda}^E)^2 \, A_\lambda 
\label{r5}
\end{eqnarray}
due to $B_\lambda^{\lambda '} = - B_{\lambda '}^\lambda$, see (\ref{biorth2b}).
$A_\lambda$ is a real number \cite{rep}.
From (\ref{r4}) and (\ref{r5}) follows
\begin{eqnarray}
\frac{\langle\hat\Psi^{E *} | 
\hat\Psi^{E} \rangle}{\la\hat\Psi^{E} | \hat\Psi^{E}\ra}
&=& \sum_\lambda (c_{\lambda}^E)^2\, A_\lambda
= \sum_\lambda \frac{(c_{\lambda}^E)^2}{ r_\lambda} \; , 
\label{r6}
\end{eqnarray}
where the definition (\ref{ph2}) for $r_\lambda$ is used. 
Then  the phase rigidity $\rho$
of the wavefunctions $\hat\Psi^E$ may be defined by 
\begin{eqnarray}
\rho &=& 
e^{2i\theta} \sum_\lambda\frac{{\rm Re}\,[(c_{\lambda}^E)^2]}{r_\lambda} 
=e^{2i\theta} \sum_\lambda \frac{1}{r_\lambda}
\bigg( [{\rm Re}(c_{\lambda}^E)]^2 - [{\rm Im}(c_{\lambda}^E)]^2 \bigg) 
\label{r7}
\end{eqnarray}
in analogy to (\ref{ph2}).
The value $\rho$ corresponds to a rotation of $\hat\Psi^E$ 
by $\theta$ corresponding to the ratio between its real and imaginary
parts.  
In spite of the complicated structure of $\rho$, it holds $1 \ge \rho \ge 0$,
see \cite{robra}. The value $\rho$ is uniquely
determined by the spectroscopic properties of the system that are expressed 
by the coupling coefficients to the environment and the level density, or 
by the  positions and widths of the resonance states and the phase 
rigidities $r_\lambda$. 
Eq. (\ref{r7}) shows the relation between $\rho$ and the $r_\lambda$.

In \cite{brscorr,robra}, the 
amplitude of the transmission through a quantum dot is considered in the
framework of  the $S$ matrix theory,
\begin{equation}
t=-2\pi
i\sum_{\lambda}\frac{\langle \xi^E_L|V|\phi_\lambda\rangle
\langle\phi_\lambda^*|V|\xi^E_R\rangle }{E-z_{\lambda}} \; . 
\label{trHeff}
\end{equation}
Here, the eigenvalues $z_\lambda$ and eigenfunctions $\phi_\lambda$ of 
$H_{\rm eff}$
are involved  with their full energy dependence, see Sect. 2.

For $\rho =1$ and well isolated resonance states, 
the transmission amplitude (\ref{trHeff}) 
repeats the resonance structure of (\ref{total}) of the wave function
$\Psi^E_C$.  The transmission peaks appear at the positions 
$E_\lambda \equiv {\rm Re}(z_\lambda)_{|E=E_\lambda} \approx E_\lambda^B$ 
of the resonance states. 
An analogous result holds  when there is a nonvanishing background term
additional to the resonance term (\ref{trHeff}) of the transmission amplitude.
The  time scale corresponding to this 
so-called {\it direct} part of the transmission is, generally, well separated
from that corresponding to the resonance part described by (\ref{trHeff})
\cite{comm1}.
Mostly, the resonances are  narrow and well separated from one another. 
They   appear as Fano resonances \cite{fano} on the smooth background.
Due to the different time scales of
the resonance and direct processes, it is $|\rho |\approx 1$ 
also in this case.  

The situation is another one when the resonances overlap.
In the overlapping regime, the resonance
states avoid crossings  with  neighbored resonance states. 
In this case   
\begin{equation}
\Gamma_\lambda <  4 \pi ~\langle
\xi^E_C|V|\phi_\lambda\rangle\langle\phi_\lambda^*|V|\xi^{E}_C\rangle   
\label{sm1o}
\end{equation}
holds even when there is only one channel in each of the two identical leads.
The relation (\ref{sm1o}) differs from  
\begin{equation}
\Gamma_\lambda =  4 \pi ~\langle
\xi^E_C|V|\phi_\lambda\rangle\langle\phi_\lambda^*|V|\xi^{E}_C\rangle   
\label{sm2o}
\end{equation}
for isolated resonances due to the 
biorthogonality of the eigenfunctions $\phi_\lambda$ 
that can not be neglected in the regime of overlapping resonances
\cite{rep}. Therefore, the contribution of the state 
$\lambda$ to $t_{(E\to E_\lambda)}$  is larger than 1. The unitarity 
condition will be fulfilled, nevertheless, due to interferences and the
possibility to  rotate the
$\phi_\lambda$, i.e. due to the non-rigidity of the phases of 
the wave functions $\phi_\lambda$. As a consequence,
the transmission in the overlapping regime does not show a resonance 
structure.  Instead, it might be nearly plateau-like, for details see
\cite{brsphas,brscorr,robra}. 
Let us rewrite therefore the transmission amplitude 
(\ref{trHeff})  by means of the  wave function (\ref{total1}),
\begin{eqnarray}
t = - 2\pi i
~\langle \xi^E_{C '}|V|\hat\Psi^E_C\rangle 
\label{tr}
\end{eqnarray}
with $\hat\Psi^E_C$ being complex, in general.
The advantage of this representation consists in the fact that it does not
suggest the existence of resonance peaks in the transmission probability. 
Quite the contrary, the transmission is determined by the degree of alignment
of the wave function $\hat \Psi_C^E$ with the propagating modes 
$\xi^E_C$ in the leads, i.e. by the phase rigidity $\rho$. 
Nevertheless, the expressions
(\ref{tr}) and (\ref{trHeff}) are fully equivalent.

The plateau-like structure of the transmission can not be  
obtained in standard quantum mechanics with fixed phases of the 
wave functions, $r_\lambda =1$ and $\rho =1$. It is generated by 
interference processes with account of the
alignment of some of the resonance states to the scattering 
states $\xi^E_C$  of the
environment. At most, ${\rm Re}\,\hat\Psi^E_C = \pm{\rm Im}\,\hat\Psi^E_C$
(in the same manner as for the $\xi^E_C$). This case corresponds to $\rho =0$. 
It will be reached when many resonance states are almost aligned with the
$\xi^E_C$, and $\sum_\lambda {\rm Re}[(c_{\lambda E})^2] /r_\lambda
\approx 0$ according to (\ref{r7}).

The numerical results \cite{brscorr} obtained by using the tight-binding
lattice Green function method \cite{datta} 
for the transmission through microwave cavities
of different shape show exactly the features discussed above. 
In the weak-coupling
regime as well as in the strong-coupling regime, the transmission shows a
resonance structure as expected from the standard quantum mechanics. 
The only difference between the two cases is the
appearance of a smooth background term in the strong-coupling regime
which does not exist in the weak-coupling case, and the reduction of the number
of resonance peaks by two (corresponding to the alignment of two resonance
states each with one channel in each of the
two identical attached leads).

In the crossover from the weak-coupling regime to the strong-coupling 
one, however, the calculated transmission 
is plateau-like instead of showing a resonance structure \cite{brscorr}. It 
is enhanced as compared to the transmission probability in the 
two borderline cases. In this regime, the resonance states overlap and 
spectroscopic reordering processes take place.
Due to widths bifurcation, some of the resonance states
become short-lived while other ones become trapped (long-lived).  
The enhancement of the transmission is caused by the short-lived states.
Most interesting is the anticorrelation between transmission $|t|$ 
and phase rigidity $|\rho|$
which can be seen  very clearly in all the numerical results
obtained in \cite{brsphas,brscorr}. The behavior of the transmission 
in the crossover
regime with overlapping resonance states does {\it not} correspond to the
expectations of the standard quantum mechanics with 
rigid phases of the  eigenfunctions of a Hermitian Hamilton operator,  and 
decay widths obtained from poles of the $S$ matrix, see e.g. \cite{shapiro}. 
Moreover, the transmission in the crossover regime
is not only enhanced but it also outspeeds the transmission
calculated in standard quantum mechanics. The reason is 
the formation of aligned (short-lived) resonance states 
in the vicinity of branch points in the complex plane.

The quantum brachistochrone problem of a physical system
can be studied by considering 
the time  needed for the transmission through the system  
from one of the attached leads to another one \cite{robra}.
The transmission time at a certain energy 
$E$ is determined by the  delay time, i.e. by the
lifetime of the resonance states lying
at this energy. The lifetime of a resonance state is bounded from below:
it can not be smaller than allowing  traveling  
through the system in accordance with traveling through the attached leads,
i.e. the system may become  transparent at most \cite{comm2}.
This lower bound can be reached in a system with the non-Hermitian 
Hamilton operator $H_{\rm eff}$  by aligning the
wave functions of the system with those of the environment
while such a possibility does not exist when $H_{\rm eff}$ is 
considered to be  Hermitian.

\section{Conclusions}

In the present paper it is shown that the FPO technique is a powerful 
method for the description of open quantum systems. The core of the method 
is the definition of two subspaces and the
non-Hermitian symmetric Hamilton operator $H_{\rm eff}$
describing the localized states in one of the two subspaces. It 
guarantees a unified description of, on the one hand, discrete and resonance
states and, on the other hand, resonance and decay processes.
The spectroscopic information is received from the eigenvalues
$z_\lambda$ and eigenfunctions $\phi_\lambda$ of $H_{\rm eff}$.
Generally the eigenvalues are complex (with the exception of BICs
and low-lying discrete states). In order to receive the spectroscopic 
information, the  two subspaces must be defined in a meaningful manner. 
Observable values that are related to the $S$ matrix and to the wave functions
$\Psi^E_C$ (being solutions of the Schr\"odinger equation with the 
Hermitian Hamilton operator $H$)
are independent of the definition of the two subspaces
as long as $P+Q=1$ is fulfilled. 
The $S$ matrix is always unitary, and it is not necessary to consider its
poles. 

The branch points in the complex plane determine the 
trajectories of the eigenvalues $z_\lambda$:  level
repulsion and widths bifurcation, respectively, appear in approaching 
them under different conditions.  This phenomenon (avoided level crossing
in the complex plane) causes the appearance of BICs 
in ${\cal P}$ symmetric systems, i.e. of ${\cal PT}$ symmetric solutions
of the Schr\"odinger equation with non-Hermitian Hamilton operator. The BICs 
are localized at all times although no selection rule forbids their decay.  
Furthermore, the phases of the $\phi_\lambda$ are not rigid in the vicinity
of the branch points. Due to this phenomenon, 
the eigenfunctions of $H_{\rm eff}$ may align to the scattering wave
functions of the environment with the consequence that
observable values are enhanced in the regime of overlapping resonances 
(where many branch points exist). 

The branch points in the complex plane influence also the properties of
discrete states: the phenomenon of avoided crossing of discrete levels 
known for a long time, can be traced back to the branch points.
Generally, the branch points in the complex plane 
introduce nonlinear effects into quantum mechanics. This fact allows, 
among others, to understand  the well-known relation 
between avoided level crossings and quantum chaos.
 
Most interesting is that,
due to the non-Hermiticity of $H_{\rm eff}$, some  problems of standard
quantum mechanics are solved without any additional assumptions. 
For example, time asymmetry appears in a natural manner since  
$H_{\rm eff}$ contains a time operator. The results obtained 
by using the FPO formalism with the non-Hermitian Hamilton operator 
$H_{\rm eff}$ may be
important for applications as well.   By controlling the
interplay between internal and external mixing of the resonance states 
by means of external parameters, the position of the localized BICs 
can be varied. Also the transmission through a system can be controlled
parametrically: a system may become transparent in the
crossover from the weak-coupling regime to the strong-coupling one.


\begin{thebibliography}{99}
\bibitem{comm2} In a 
many-particle system, the life time of a resonance state can not
be smaller than the lifetime of a single-particle resonance with the
corresponding quantum numbers.
\end{thebibliography}
\end{document}